
%
%
%
%
%
%
%
%
\def\standardrisposta{s }\def\reducedrisposta{r }
\def\mplarisposta{mpla }\def\zerorisposta{z }
\def\doublerisposta{d }\def\cartarisposta{e }\def\amsrisposta{y }
\newcount\ingrandimento \newcount\sinnota \newcount\dimnota
\newcount\unoduecol \newdimen\collhsize \newdimen\tothsize
\newdimen\fullhsize \newcount\controllorisposta \sinnota=1
\newskip\infralinea  \global\controllorisposta=0
\immediate\write16 { ********  Welcome to PANDA macros (Plain TeX,
AP, 1991) ******** }
\immediate\write16 { You'll have to answer a few questions in
lowercase.}
\message{>  Do you want it in double-page (d), reduced (r)
or standard format (s) ? }\read-1 to\risposta
\message{>  Do you want it in USA A4 (u) or EUROPEAN A4
(e) paper size ? }\read-1 to\srisposta
\message{>  Do you have AMSFonts 2.0 (math) fonts (y/n) ? }
\read-1 to\arisposta
%
%
%
%
%
\ifx\risposta\standardrisposta \ingrandimento=1200
\message {>> This will come out UNREDUCED << }
\dimnota=2 \unoduecol=1 \global\controllorisposta=1 \fi
\ifx\risposta\reducedrisposta \ingrandimento=1095 \dimnota=1
\unoduecol=1  \global\controllorisposta=1
\message {>> This will come out REDUCED << } \fi
\ifx\risposta\doublerisposta \ingrandimento=1000 \dimnota=2
\unoduecol=2   \message {>> You must print this in
LANDSCAPE orientation << } \global\controllorisposta=1 \fi
\ifx\risposta\mplarisposta \ingrandimento=1000 \dimnota=1
\message {>> Mod. Phys. Lett. A format << }
\unoduecol=1 \global\controllorisposta=1 \fi
\ifx\risposta\zerorisposta \ingrandimento=1000 \dimnota=2
\message {>> Zero Magnification format << }
\unoduecol=1 \global\controllorisposta=1 \fi
\ifnum\controllorisposta=0  \ingrandimento=1200
\message {>>> ERROR IN INPUT, I ASSUME STANDARD
UNREDUCED FORMAT <<< }  \dimnota=2 \unoduecol=1 \fi
\magnification=\ingrandimento
%
%
%
%
\newdimen\eucolumnsize \newdimen\eudoublehsize \newdimen\eudoublevsize
\newdimen\uscolumnsize \newdimen\usdoublehsize \newdimen\usdoublevsize
\newdimen\eusinglehsize \newdimen\eusinglevsize \newdimen\ussinglehsize
\newskip\standardbaselineskip \newdimen\ussinglevsize
\newskip\reducedbaselineskip \newskip\doublebaselineskip
\eucolumnsize=12.0truecm    
\eudoublehsize=25.5truecm   
\eudoublevsize=6.5truein    
\uscolumnsize=4.4truein     
\usdoublehsize=9.4truein    
\usdoublevsize=6.8truein    
\eusinglehsize=6.5truein    
\eusinglevsize=24truecm     
\ussinglehsize=6.5truein    
\ussinglevsize=8.9truein    
\standardbaselineskip=16pt plus.2pt  
\reducedbaselineskip=14pt plus.2pt   
\doublebaselineskip=12pt plus.2pt    
%
%
\def\Portoffset{}
\def\Landoffset{}
\ifx\risposta\mplarisposta \def\Portoffset{\hoffset=1.8truecm} \fi
%
%
\def\Landspec{}
\tolerance=10000
\parskip=0pt plus2pt  \leftskip=0pt \rightskip=0pt
%
%
\ifx\risposta\standardrisposta \infralinea=\standardbaselineskip \fi
\ifx\risposta\reducedrisposta  \infralinea=\reducedbaselineskip \fi
\ifx\risposta\doublerisposta   \infralinea=\doublebaselineskip \fi
\ifx\risposta\mplarisposta     \infralinea=13pt \fi
\ifx\risposta\zerorisposta     \infralinea=12pt plus.2pt\fi
\ifnum\controllorisposta=0    \infralinea=\standardbaselineskip \fi
\ifx\risposta\doublerisposta   \Landoffset \else \Portoffset \fi
\ifx\risposta\doublerisposta \ifx\srisposta\cartarisposta
\tothsize=\eudoublehsize \collhsize=\eucolumnsize
\vsize=\eudoublevsize  \else  \tothsize=\usdoublehsize
\collhsize=\uscolumnsize \vsize=\usdoublevsize \fi \else
\ifx\srisposta\cartarisposta \tothsize=\eusinglehsize
\vsize=\eusinglevsize \else  \tothsize=\ussinglehsize
\vsize=\ussinglevsize \fi \collhsize=4.4truein \fi
\ifx\risposta\mplarisposta \tothsize=5.0truein
\vsize=7.8truein \collhsize=4.4truein \fi
%
%
%
%
\newcount\contaeuler \newcount\contacyrill \newcount\contaams
\font\ninerm=cmr9  \font\eightrm=cmr8  \font\sixrm=cmr6
\font\ninei=cmmi9  \font\eighti=cmmi8  \font\sixi=cmmi6
\font\ninesy=cmsy9  \font\eightsy=cmsy8  \font\sixsy=cmsy6
\font\ninebf=cmbx9  \font\eightbf=cmbx8  \font\sixbf=cmbx6
\font\ninett=cmtt9  \font\eighttt=cmtt8  \font\nineit=cmti9
\font\eightit=cmti8 \font\ninesl=cmsl9  \font\eightsl=cmsl8
\skewchar\ninei='177 \skewchar\eighti='177 \skewchar\sixi='177
\skewchar\ninesy='60 \skewchar\eightsy='60 \skewchar\sixsy='60
\hyphenchar\ninett=-1 \hyphenchar\eighttt=-1 \hyphenchar\tentt=-1
\def\bfmath{\cmmib}                 
\font\tencmmib=cmmib10  \newfam\cmmibfam  \skewchar\tencmmib='177
\font\tencmbsy=cmbsy10  \newfam\cmbsyfam  \skewchar\tencmbsy='60
\def\scaps{\cmcsc}                 
\font\tencmcsc=cmcsc10  \newfam\cmcscfam
\ifnum\ingrandimento=1095

\font\capsone=cmcsc10 at 10.95pt 

\else

\font\capsone=cmcsc10 at 12pt 
\fi

\def\ttaarr{\bf}		
\def\ppaarr{\sl}		

%
%
%
\newfam\eufmfam \newfam\msamfam \newfam\msbmfam \newfam\eufbfam
\def\Loadeulerfonts{\global\contaeuler=1 \ifx\arisposta\amsrisposta
\font\teneufm=eufm10              
\font\eighteufm=eufm8 \font\nineeufm=eufm9 \font\sixeufm=eufm6
\font\seveneufm=eufm7  \font\fiveeufm=eufm5
\font\teneufb=eufb10              
\font\eighteufb=eufb8 \font\nineeufb=eufb9 \font\sixeufb=eufb6
\font\seveneufb=eufb7  \font\fiveeufb=eufb5
\font\teneurm=eurm10              
\font\eighteurm=eurm8 \font\nineeurm=eurm9
\font\teneurb=eurb10              
\font\eighteurb=eurb8 \font\nineeurb=eurb9
\font\teneusm=eusm10              
\font\eighteusm=eusm8 \font\nineeusm=eusm9
\font\teneusb=eusb10              
\font\eighteusb=eusb8 \font\nineeusb=eusb9
\else \def\eufm{\tt} \def\eufb{\tt} \def\eurm{\tt} \def\eurb{\tt}
\def\eusm{\tt} \def\eusb{\tt}    \fi}
\def\loadeuler{\Loadeulerfonts\tenpoint}
\def\loadamsmath{\global\contaams=1 \ifx\arisposta\amsrisposta
\font\tenmsam=msam10 \font\ninemsam=msam9 \font\eightmsam=msam8
\font\sevenmsam=msam7 \font\sixmsam=msam6 \font\fivemsam=msam5
\font\tenmsbm=msbm10 \font\ninemsbm=msbm9 \font\eightmsbm=msbm8
\font\sevenmsbm=msbm7 \font\sixmsbm=msbm6 \font\fivemsbm=msbm5
\else \def\msbm{\bf} \fi \def\Bbb{\msbm} \def\symbl{\msam} \tenpoint}
\def\loadcyrill{\global\contacyrill=1 \ifx\arisposta\amsrisposta
\font\tenwncyr=wncyr10 \font\ninewncyr=wncyr9 \font\eightwncyr=wncyr8
\font\tenwncyb=wncyr10 \font\ninewncyb=wncyr9 \font\eightwncyb=wncyr8
\font\tenwncyi=wncyr10 \font\ninewncyi=wncyr9 \font\eightwncyi=wncyr8
\else \def\cyrill{\sl} \def\cyrilb{\sl} \def\cyrili{\sl} \fi\tenpoint}
\ifx\arisposta\amsrisposta
\font\sevenex=cmex7               
\font\eightex=cmex8  \font\nineex=cmex9
\font\ninecmmib=cmmib9   \font\eightcmmib=cmmib8
\font\sevencmmib=cmmib7 \font\sixcmmib=cmmib6
\font\fivecmmib=cmmib5   \skewchar\ninecmmib='177
\skewchar\eightcmmib='177  \skewchar\sevencmmib='177
\skewchar\sixcmmib='177   \skewchar\fivecmmib='177
%
%
\font\ninecmbsy=cmbsy9    \font\eightcmbsy=cmbsy8
\font\sevencmbsy=cmbsy7  \font\sixcmbsy=cmbsy6
\font\fivecmbsy=cmbsy5   \skewchar\ninecmbsy='60
\skewchar\eightcmbsy='60  \skewchar\sevencmbsy='60
\skewchar\sixcmbsy='60    \skewchar\fivecmbsy='60
\font\ninecmcsc=cmcsc9    \font\eightcmcsc=cmcsc8     \else
\def\cmmib{\fam\cmmibfam\tencmmib}\textfont\cmmibfam=\tencmmib
\scriptfont\cmmibfam=\tencmmib \scriptscriptfont\cmmibfam=\tencmmib
\def\cmbsy{\fam\cmbsyfam\tencmbsy} \textfont\cmbsyfam=\tencmbsy
\scriptfont\cmbsyfam=\tencmbsy \scriptscriptfont\cmbsyfam=\tencmbsy
\scriptfont\cmcscfam=\tencmcsc \scriptscriptfont\cmcscfam=\tencmcsc
\def\cmcsc{\fam\cmcscfam\tencmcsc} \textfont\cmcscfam=\tencmcsc \fi
\catcode`@=11
\newskip\ttglue
\gdef\tenpoint{\def\rm{\fam0\tenrm}
  \textfont0=\tenrm \scriptfont0=\sevenrm \scriptscriptfont0=\fiverm
  \textfont1=\teni \scriptfont1=\seveni \scriptscriptfont1=\fivei
  \textfont2=\tensy \scriptfont2=\sevensy \scriptscriptfont2=\fivesy
  \textfont3=\tenex \scriptfont3=\tenex \scriptscriptfont3=\tenex
  \def\mcal{\fam2 \tensy}  \def\mmit{\fam1 \teni}
  \textfont\itfam=\tenit \def\it{\fam\itfam\tenit}
  \textfont\slfam=\tensl \def\sl{\fam\slfam\tensl}
  \textfont\ttfam=\tentt \scriptfont\ttfam=\eighttt
  \scriptscriptfont\ttfam=\eighttt  \def\tt{\fam\ttfam\tentt}
  \textfont\bffam=\tenbf \scriptfont\bffam=\sevenbf
  \scriptscriptfont\bffam=\fivebf \def\bf{\fam\bffam\tenbf}
     \ifx\arisposta\amsrisposta    \ifnum\contaeuler=1
  \textfont\eufmfam=\teneufm \scriptfont\eufmfam=\seveneufm
  \scriptscriptfont\eufmfam=\fiveeufm \def\eufm{\fam\eufmfam\teneufm}
  \textfont\eufbfam=\teneufb \scriptfont\eufbfam=\seveneufb
  \scriptscriptfont\eufbfam=\fiveeufb \def\eufb{\fam\eufbfam\teneufb}
  \def\eurm{\teneurm} \def\eurb{\teneurb} \def\eusm{\teneusm}
  \def\eusb{\teneusb}    \fi    \ifnum\contaams=1
  \textfont\msamfam=\tenmsam \scriptfont\msamfam=\sevenmsam
  \scriptscriptfont\msamfam=\fivemsam \def\msam{\fam\msamfam\tenmsam}
  \textfont\msbmfam=\tenmsbm \scriptfont\msbmfam=\sevenmsbm
  \scriptscriptfont\msbmfam=\fivemsbm \def\msbm{\fam\msbmfam\tenmsbm}
     \fi      \ifnum\contacyrill=1     \def\cyrill{\tenwncyr}
  \def\cyrilb{\tenwncyb}  \def\cyrili{\tenwncyi}         \fi
  \textfont3=\tenex \scriptfont3=\sevenex \scriptscriptfont3=\sevenex
  \def\cmmib{\fam\cmmibfam\tencmmib} \scriptfont\cmmibfam=\sevencmmib
  \textfont\cmmibfam=\tencmmib  \scriptscriptfont\cmmibfam=\fivecmmib
  \def\cmbsy{\fam\cmbsyfam\tencmbsy} \scriptfont\cmbsyfam=\sevencmbsy
  \textfont\cmbsyfam=\tencmbsy  \scriptscriptfont\cmbsyfam=\fivecmbsy
  \def\cmcsc{\fam\cmcscfam\tencmcsc} \scriptfont\cmcscfam=\eightcmcsc
  \textfont\cmcscfam=\tencmcsc \scriptscriptfont\cmcscfam=\eightcmcsc
     \fi            \tt \ttglue=.5em plus.25em minus.15em
  \normalbaselineskip=12pt
  \setbox\strutbox=\hbox{\vrule height8.5pt depth3.5pt width0pt}
  \let\sc=\eightrm \let\big=\tenbig   \normalbaselines
  \baselineskip=\infralinea  \rm}
\gdef\ninepoint{\def\rm{\fam0\ninerm}
  \textfont0=\ninerm \scriptfont0=\sixrm \scriptscriptfont0=\fiverm
  \textfont1=\ninei \scriptfont1=\sixi \scriptscriptfont1=\fivei
  \textfont2=\ninesy \scriptfont2=\sixsy \scriptscriptfont2=\fivesy
  \textfont3=\tenex \scriptfont3=\tenex \scriptscriptfont3=\tenex
  \def\mcal{\fam2 \ninesy}  \def\mmit{\fam1 \ninei}
  \textfont\itfam=\nineit \def\it{\fam\itfam\nineit}
  \textfont\slfam=\ninesl \def\sl{\fam\slfam\ninesl}
  \textfont\ttfam=\ninett \scriptfont\ttfam=\eighttt
  \scriptscriptfont\ttfam=\eighttt \def\tt{\fam\ttfam\ninett}
  \textfont\bffam=\ninebf \scriptfont\bffam=\sixbf
  \scriptscriptfont\bffam=\fivebf \def\bf{\fam\bffam\ninebf}
     \ifx\arisposta\amsrisposta  \ifnum\contaeuler=1
  \textfont\eufmfam=\nineeufm \scriptfont\eufmfam=\sixeufm
  \scriptscriptfont\eufmfam=\fiveeufm \def\eufm{\fam\eufmfam\nineeufm}
  \textfont\eufbfam=\nineeufb \scriptfont\eufbfam=\sixeufb
  \scriptscriptfont\eufbfam=\fiveeufb \def\eufb{\fam\eufbfam\nineeufb}
  \def\eurm{\nineeurm} \def\eurb{\nineeurb} \def\eusm{\nineeusm}
  \def\eusb{\nineeusb}     \fi   \ifnum\contaams=1
  \textfont\msamfam=\ninemsam \scriptfont\msamfam=\sixmsam
  \scriptscriptfont\msamfam=\fivemsam \def\msam{\fam\msamfam\ninemsam}
  \textfont\msbmfam=\ninemsbm \scriptfont\msbmfam=\sixmsbm
  \scriptscriptfont\msbmfam=\fivemsbm \def\msbm{\fam\msbmfam\ninemsbm}
     \fi       \ifnum\contacyrill=1     \def\cyrill{\ninewncyr}
  \def\cyrilb{\ninewncyb}  \def\cyrili{\ninewncyi}         \fi
  \textfont3=\nineex \scriptfont3=\sevenex \scriptscriptfont3=\sevenex
  \def\cmmib{\fam\cmmibfam\ninecmmib}  \textfont\cmmibfam=\ninecmmib
  \scriptfont\cmmibfam=\sixcmmib \scriptscriptfont\cmmibfam=\fivecmmib
  \def\cmbsy{\fam\cmbsyfam\ninecmbsy}  \textfont\cmbsyfam=\ninecmbsy
  \scriptfont\cmbsyfam=\sixcmbsy \scriptscriptfont\cmbsyfam=\fivecmbsy
  \def\cmcsc{\fam\cmcscfam\ninecmcsc} \scriptfont\cmcscfam=\eightcmcsc
  \textfont\cmcscfam=\ninecmcsc \scriptscriptfont\cmcscfam=\eightcmcsc
     \fi            \tt \ttglue=.5em plus.25em minus.15em
  \normalbaselineskip=11pt
  \setbox\strutbox=\hbox{\vrule height8pt depth3pt width0pt}
  \let\sc=\sevenrm \let\big=\ninebig \normalbaselines\rm}
\gdef\eightpoint{\def\rm{\fam0\eightrm}
  \textfont0=\eightrm \scriptfont0=\sixrm \scriptscriptfont0=\fiverm
  \textfont1=\eighti \scriptfont1=\sixi \scriptscriptfont1=\fivei
  \textfont2=\eightsy \scriptfont2=\sixsy \scriptscriptfont2=\fivesy
  \textfont3=\tenex \scriptfont3=\tenex \scriptscriptfont3=\tenex
  \def\mcal{\fam2 \eightsy}  \def\mmit{\fam1 \eighti}
  \textfont\itfam=\eightit \def\it{\fam\itfam\eightit}
  \textfont\slfam=\eightsl \def\sl{\fam\slfam\eightsl}
  \textfont\ttfam=\eighttt \scriptfont\ttfam=\eighttt
  \scriptscriptfont\ttfam=\eighttt \def\tt{\fam\ttfam\eighttt}
  \textfont\bffam=\eightbf \scriptfont\bffam=\sixbf
  \scriptscriptfont\bffam=\fivebf \def\bf{\fam\bffam\eightbf}
     \ifx\arisposta\amsrisposta   \ifnum\contaeuler=1
  \textfont\eufmfam=\eighteufm \scriptfont\eufmfam=\sixeufm
  \scriptscriptfont\eufmfam=\fiveeufm \def\eufm{\fam\eufmfam\eighteufm}
  \textfont\eufbfam=\eighteufb \scriptfont\eufbfam=\sixeufb
  \scriptscriptfont\eufbfam=\fiveeufb \def\eufb{\fam\eufbfam\eighteufb}
  \def\eurm{\eighteurm} \def\eurb{\eighteurb} \def\eusm{\eighteusm}
  \def\eusb{\eighteusb}       \fi    \ifnum\contaams=1
  \textfont\msamfam=\eightmsam \scriptfont\msamfam=\sixmsam
  \scriptscriptfont\msamfam=\fivemsam \def\msam{\fam\msamfam\eightmsam}
  \textfont\msbmfam=\eightmsbm \scriptfont\msbmfam=\sixmsbm
  \scriptscriptfont\msbmfam=\fivemsbm \def\msbm{\fam\msbmfam\eightmsbm}
     \fi       \ifnum\contacyrill=1     \def\cyrill{\eightwncyr}
  \def\cyrilb{\eightwncyb}  \def\cyrili{\eightwncyi}         \fi
  \textfont3=\eightex \scriptfont3=\sevenex \scriptscriptfont3=\sevenex
  \def\cmmib{\fam\cmmibfam\eightcmmib}  \textfont\cmmibfam=\eightcmmib
  \scriptfont\cmmibfam=\sixcmmib \scriptscriptfont\cmmibfam=\fivecmmib
  \def\cmbsy{\fam\cmbsyfam\eightcmbsy}  \textfont\cmbsyfam=\eightcmbsy
  \scriptfont\cmbsyfam=\sixcmbsy \scriptscriptfont\cmbsyfam=\fivecmbsy
  \def\cmcsc{\fam\cmcscfam\eightcmcsc} \scriptfont\cmcscfam=\eightcmcsc
  \textfont\cmcscfam=\eightcmcsc \scriptscriptfont\cmcscfam=\eightcmcsc
     \fi             \tt \ttglue=.5em plus.25em minus.15em
  \normalbaselineskip=9pt
  \setbox\strutbox=\hbox{\vrule height7pt depth2pt width0pt}
  \let\sc=\sixrm \let\big=\eightbig \normalbaselines\rm }
\gdef\tenbig#1{{\hbox{$\left#1\vbox to8.5pt{}\right.\n@space$}}}
\gdef\ninebig#1{{\hbox{$\textfont0=\tenrm\textfont2=\tensy
   \left#1\vbox to7.25pt{}\right.\n@space$}}}
\gdef\eightbig#1{{\hbox{$\textfont0=\ninerm\textfont2=\ninesy
   \left#1\vbox to6.5pt{}\right.\n@space$}}}
\def\alternativefont#1#2{\ifx\arisposta\amsrisposta \relax \else
\xdef#1{#2} \fi}
\global\contaeuler=0 \global\contacyrill=0 \global\contaams=0
%
%
%
%
\newbox\fotlinebb \newbox\hedlinebb \newbox\leftcolumn
\gdef\makeheadline{\vbox to 0pt{\vskip-22.5pt
     \fullline{\vbox to8.5pt{}\the\headline}\vss}\nointerlineskip}
\gdef\makehedlinebb{\vbox to 0pt{\vskip-22.5pt
     \fullline{\vbox to8.5pt{}\copy\hedlinebb\hfil
     \line{\hfill\the\headline\hfill}}\vss} \nointerlineskip}
\gdef\makefootline{\baselineskip=24pt \fullline{\the\footline}}
\gdef\makefotlinebb{\baselineskip=24pt
    \fullline{\copy\fotlinebb\hfil\line{\hfill\the\footline\hfill}}}
\gdef\doubleformat{\shipout\vbox{\Landspec\makehedlinebb
     \fullline{\box\leftcolumn\hfil\columnbox}\makefotlinebb}
     \advancepageno}
\gdef\columnbox{\leftline{\pagebody}}
\gdef\line#1{\hbox to\hsize{\hskip\leftskip#1\hskip\rightskip}}
\gdef\fullline#1{\hbox to\fullhsize{\hskip\leftskip{#1}%
\hskip\rightskip}}
\gdef\footnote#1{\let\@sf=\empty
         \ifhmode\edef\#sf{\spacefactor=\the\spacefactor}\/\fi
         #1\@sf\vfootnote{#1}}
\gdef\vfootnote#1{\insert\footins\bgroup
         \ifnum\dimnota=1  \eightpoint\fi
         \ifnum\dimnota=2  \ninepoint\fi
         \ifnum\dimnota=0  \tenpoint\fi
         \interlinepenalty=\interfootnotelinepenalty
         \splittopskip=\ht\strutbox
         \splitmaxdepth=\dp\strutbox \floatingpenalty=20000
         \leftskip=\oldssposta \rightskip=\olddsposta
         \spaceskip=0pt \xspaceskip=0pt
         \ifnum\sinnota=0   \textindent{#1}\fi
         \ifnum\sinnota=1   \item{#1}\fi
         \footstrut\futurelet\next\fo@t}
\gdef\fo@t{\ifcat\bgroup\noexpand\next \let\next\f@@t
             \else\let\next\f@t\fi \next}
\gdef\f@@t{\bgroup\aftergroup\@foot\let\next}
\gdef\f@t#1{#1\@foot} \gdef\@foot{\strut\egroup}
\gdef\footstrut{\vbox to\splittopskip{}}
\skip\footins=\bigskipamount
\count\footins=1000  \dimen\footins=8in
\catcode`@=12
\tenpoint
\ifnum\unoduecol=1 \hsize=\tothsize   \fullhsize=\tothsize \fi
\ifnum\unoduecol=2 \hsize=\collhsize  \fullhsize=\tothsize \fi
\global\let\lrcol=L      \ifnum\unoduecol=1
\output{\plainoutput{\ifnum\tipbnota=2 \clearnmbnota\fi}} \fi
\ifnum\unoduecol=2 \output{\if L\lrcol
     \global\setbox\leftcolumn=\columnbox
     \global\setbox\fotlinebb=\line{\hfill\the\footline\hfill}
     \global\setbox\hedlinebb=\line{\hfill\the\headline\hfill}
     \advancepageno  \global\let\lrcol=R
     \else  \doubleformat \global\let\lrcol=L \fi
     \ifnum\outputpenalty>-20000 \else\dosupereject\fi
     \ifnum\tipbnota=2\clearnmbnota\fi }\fi
\def\ifdoublepage{\ifnum\unoduecol=2 }
\gdef\yespagenumbers{\footline={\hss\tenrm\folio\hss}}
\gdef\ciao{ \ifnum\fdefcontre=1 \endfdef\fi
     \par\vfill\supereject \ifnum\unoduecol=2
     \if R\lrcol  \headline={}\nopagenumbers\null\vfill\eject
     \fi\fi \end}

\newskip\olddsposta \newskip\oldssposta
\global\oldssposta=\leftskip \global\olddsposta=\rightskip

\def\filldots{\leaders\hbox to 1em{\hss.\hss}\hfill}
\def\inquadrb#1 {\vbox {\hrule  \hbox{\vrule \vbox {\vskip .2cm
    \hbox {\ #1\ } \vskip .2cm } \vrule  }  \hrule} }
 \def\newline{\hfil\break}
\def\jump{\vskip\baselineskip} \newskip\iinnffrr
\def\sjump{\iinnffrr=\baselineskip
          \divide\iinnffrr by 2 \vskip\iinnffrr}
\def\bjump{\vskip\baselineskip \vskip\baselineskip}
\newcount\nmbnota  \def\clearnmbnota{\global\nmbnota=0}
\newcount\tipbnota \def\letterfootnote{\global\tipbnota=1}

\def\note#1{\global\advance\nmbnota by 1 \ifnum\tipbnota=1
    \footnote{$^{\rm\nttlett}$}{#1} \else {\ifnum\tipbnota=2
    \footnote{$^{\nttsymb}$}{#1}
    \else\footnote{$^{\the\nmbnota}$}{#1}\fi}\fi}
\def\nttlett{\ifcase\nmbnota \or a\or b\or c\or d\or e\or f\or
g\or h\or i\or j\or k\or l\or m\or n\or o\or p\or q\or r\or
s\or t\or u\or v\or w\or y\or x\or z\fi}
\def\nttsymb{\ifcase\nmbnota \or\dag\or\sharp\or\ddag\or\star\or
\natural\or\flat\or\clubsuit\or\diamondsuit\or\heartsuit
\or\spadesuit\fi}   \clearnmbnota
\def\numberfootnote{\global\tipbnota=0} \numberfootnote
\def\setnote#1{\expandafter\xdef\csname#1\endcsname{
\ifnum\tipbnota=1 {\rm\nttlett} \else {\ifnum\tipbnota=2
{\nttsymb} \else \the\nmbnota\fi}\fi} }
\newcount\nbmfig  \def\clearnbmfig{\global\nbmfig=0}
\gdef\figure{\global\advance\nbmfig by 1
      {\rm fig. \the\nbmfig}}   \clearnbmfig
\def\setfig#1{\expandafter\xdef\csname#1\endcsname{fig. \the\nbmfig}}
 \def\endformula{\eqno\numero $$}
 \def\efr{\endformula}
\newcount\frmcount \def\clearfrmcount{\global\frmcount=0}
\def\numero{\global\advance\frmcount by 1   \ifnum\indappcount=0
  {\ifnum\cpcount <1 {\hbox{\rm (\the\frmcount )}}  \else
  {\hbox{\rm (\the\cpcount .\the\frmcount )}} \fi}  \else
  {\hbox{\rm (\applett .\the\frmcount )}} \fi}
\def\nameformula#1{\global\advance\frmcount by 1%
\ifnum\draftnum=0  {\ifnum\indappcount=0%
{\ifnum\cpcount<1\xdef\spzzttrra{(\the\frmcount )}%
\else\xdef\spzzttrra{(\the\cpcount .\the\frmcount )}\fi}%
\else\xdef\spzzttrra{(\applett .\the\frmcount )}\fi}%
\else\xdef\spzzttrra{(#1)}\fi%
\expandafter\xdef\csname#1\endcsname{\spzzttrra}
\eqno \hbox{\rm\spzzttrra} $$}
\def\nfr{\nameformula}    
\def\nameali#1{\global\advance\frmcount by 1%
\ifnum\draftnum=0  {\ifnum\indappcount=0%
{\ifnum\cpcount<1\xdef\spzzttrra{(\the\frmcount )}%
\else\xdef\spzzttrra{(\the\cpcount .\the\frmcount )}\fi}%
\else\xdef\spzzttrra{(\applett .\the\frmcount )}\fi}%
\else\xdef\spzzttrra{(#1)}\fi%
\expandafter\xdef\csname#1\endcsname{\spzzttrra}
  \hbox{\rm\spzzttrra} }      \clearfrmcount
\newcount\cpcount \def\clearcpcount{\global\cpcount=0}
\newcount\subcpcount \def\clearsubcpcount{\global\subcpcount=0}
\newcount\appcount \def\clearappcount{\global\appcount=0}
\newcount\indappcount \def\clearindappcount{\indappcount=0}
\newcount\sottoparcount 

\def\applett{\ifcase\appcount  \or {A}\or {B}\or {C}\or
{D}\or {E}\or {F}\or {G}\or {H}\or {I}\or {J}\or {K}\or {L}\or
{M}\or {N}\or {O}\or {P}\or {Q}\or {R}\or {S}\or {T}\or {U}\or
{V}\or {W}\or {X}\or {Y}\or {Z}\fi    \ifnum\appcount<0
\immediate\write16 {Panda ERROR - Appendix: counter "appcount"
out of range}\fi  \ifnum\appcount>26  \immediate\write16 {Panda
ERROR - Appendix: counter "appcount" out of range}\fi}
\clearappcount  \clearindappcount \newcount\connttrre
\def\clearconnttrre{\global\connttrre=0} \newcount\countref
\def\clearcountref{\global\countref=0} \clearcountref
\def\chapter#1{\global\advance\cpcount by 1 \clearfrmcount
                 \goodbreak\null\vbox{\jump\nobreak
                 \clearsubcpcount\clearindappcount
                 \itemitem{\ttaarr\the\cpcount .\qquad}{\ttaarr #1}
                 \par\nobreak\jump\sjump}\nobreak}
\def\section#1{\global\advance\subcpcount by 1 \goodbreak\null
               \vbox{\sjump\nobreak\ifnum\indappcount=0
                 {\ifnum\cpcount=0 {\itemitem{\ppaarr
               .\the\subcpcount\quad\enskip\ }{\ppaarr #1}\par} \else
                 {\itemitem{\ppaarr\the\cpcount .\the\subcpcount\quad
                  \enskip\ }{\ppaarr #1} \par}  \fi}
                \else{\itemitem{\ppaarr\applett .\the\subcpcount\quad
                 \enskip\ }{\ppaarr #1}\par}\fi\nobreak\jump}\nobreak}
\clearsubcpcount
\def\appendix#1{\global\advance\appcount by 1 \clearfrmcount
                  \goodbreak\null\vbox{\jump\nobreak
                  \global\advance\indappcount by 1 \clearsubcpcount
          \itemitem{ }{\hskip-40pt\ttaarr Appendix\ \applett :\ #1}
             \nobreak\jump\sjump}\nobreak}
\clearappcount \clearindappcount
\def\references{\goodbreak\null\vbox{\jump\nobreak
   \itemitem{}{\ttaarr References} \nobreak\jump\sjump}\nobreak}

\clearcpcount\clearcountref

\def\setchap#1{\ifnum\indappcount=0{\ifnum\subcpcount=0%
\xdef\spzzttrra{\the\cpcount}%
\else\xdef\spzzttrra{\the\cpcount .\the\subcpcount}\fi}
\else{\ifnum\subcpcount=0 \xdef\spzzttrra{\applett}%
\else\xdef\spzzttrra{\applett .\the\subcpcount}\fi}\fi
\expandafter\xdef\csname#1\endcsname{\spzzttrra}}
\newcount\draftnum \newcount\ppora   \newcount\ppminuti
\global\ppora=\time   \global\ppminuti=\time
\global\divide\ppora by 60  \draftnum=\ppora
\multiply\draftnum by 60    \global\advance\ppminuti by -\draftnum
\def\droggi{\number\day /\number\month /\number\year\ \the\ppora
:\the\ppminuti}     \global\draftnum=0
\def\draftcomment#1{\ifnum\draftnum=0 \relax \else
{\ {\bf ***}\ #1\ {\bf ***}\ }\fi} 
%
%
\catcode`@=11
\gdef\Ref#1{\expandafter\ifx\csname @rrxx@#1\endcsname\relax%
{\global\advance\countref by 1    \ifnum\countref>200
\immediate\write16 {Panda ERROR - Ref: maximum number of references
exceeded}  \expandafter\xdef\csname @rrxx@#1\endcsname{0}\else
\expandafter\xdef\csname @rrxx@#1\endcsname{\the\countref}\fi}\fi
\ifnum\draftnum=0 \csname @rrxx@#1\endcsname \else#1\fi}
\gdef\beginref{\ifnum\draftnum=0  \gdef\Rref{\fairef}
\gdef\endref{\scriviref} \else\relax\fi
\ifx\risposta\mplarisposta \ninepoint \fi
\parskip 2pt plus.2pt \baselineskip=12pt}
\def\Reflab#1{[#1]} \gdef\Rref#1#2{\item{\Reflab{#1}}{#2}}
\gdef\endref{\relax}  \newcount\conttemp
\gdef\fairef#1#2{\expandafter\ifx\csname @rrxx@#1\endcsname\relax
{\global\conttemp=0 \immediate\write16 {Panda ERROR - Ref: reference
[#1] undefined}} \else
{\global\conttemp=\csname @rrxx@#1\endcsname } \fi
\global\advance\conttemp by 50  \global\setbox\conttemp=\hbox{#2} }
\gdef\scriviref{\clearconnttrre\conttemp=50
\loop\ifnum\connttrre<\countref \advance\conttemp by 1
\advance\connttrre by 1
\item{\Reflab{\the\connttrre}}{\unhcopy\conttemp} \repeat}
\clearcountref \clearconnttrre
\catcode`@=12
\ifx\risposta\mplarisposta \def\Reflab#1{#1.} \letterfootnote \fi

\def\slashchar#1{\setbox0=\hbox{$#1$} \dimen0=\wd0
     \setbox1=\hbox{/} \dimen1=\wd1 \ifdim\dimen0>\dimen1
      \rlap{\hbox to \dimen0{\hfil/\hfil}} #1 \else
      \rlap{\hbox to \dimen1{\hfil$#1$\hfil}} / \fi}
\ifx\oldchi\undefined \let\oldchi=\chi
  \def\cchi{{\raise 1pt\hbox{$\oldchi$}}} \let\chi=\cchi \fi

\def\frac#1#2{{\textstyle{#1 \over #2}}}

\def\half{\ifinner {\scriptstyle {1 \over 2}}\else {1 \over 2} \fi}

\def\simge{\rlap{\raise 2pt \hbox{$>$}}{\lower 2pt \hbox{$\sim$}}}
\def\simle{\rlap{\raise 2pt \hbox{$<$}}{\lower 2pt \hbox{$\sim$}}}

\def\vbig#1#2{{\vbigd@men=#2\divide\vbigd@men by 2%
\hbox{$\left#1\vbox to \vbigd@men{}\right.\n@space$}}}

%
%
\newcount\fdefcontre \newcount\fdefcount \newcount\indcount
\newread\filefdef  \newread\fileftmp  \newwrite\filefdef
\newwrite\fileftmp     \def\strip#1*.A {#1}
\def\futuredef#1{\beginfdef
\expandafter\ifx\csname#1\endcsname\relax%
{\immediate\write\fileftmp {#1*.A}
\immediate\write16 {Panda Warning - fdef: macro "#1" on page
\the\pageno \space undefined}
\ifnum\draftnum=0 \expandafter\xdef\csname#1\endcsname{(?)}
\else \expandafter\xdef\csname#1\endcsname{(#1)} \fi
\global\advance\fdefcount by 1}\fi   \csname#1\endcsname}

\def\beginfdef{\ifnum\fdefcontre=0
\immediate\openin\filefdef \jobname.fdef
\immediate\openout\fileftmp \jobname.ftmp
\global\fdefcontre=1  \ifeof\filefdef \immediate\write16 {Panda
WARNING - fdef: file \jobname.fdef not found, run TeX again}
\else \immediate\read\filefdef to\spzzttrra
\global\advance\fdefcount by \spzzttrra
\indcount=0      \loop\ifnum\indcount<\fdefcount
\advance\indcount by 1   \immediate\read\filefdef to\spezttrra
\immediate\read\filefdef to\sppzttrra
\edef\spzzttrra{\expandafter\strip\spezttrra}
\immediate\write\fileftmp {\spzzttrra *.A}
\expandafter\xdef\csname\spzzttrra\endcsname{\sppzttrra}
\repeat \fi \immediate\closein\filefdef \fi}
\def\endfdef{\immediate\closeout\fileftmp   \ifnum\fdefcount>0
\immediate\openin\fileftmp \jobname.ftmp
\immediate\openout\filefdef \jobname.fdef
\immediate\write\filefdef {\the\fdefcount}   \indcount=0
\loop\ifnum\indcount<\fdefcount    \advance\indcount by 1
\immediate\read\fileftmp to\spezttrra
\edef\spzzttrra{\expandafter\strip\spezttrra}
\immediate\write\filefdef{\spzzttrra *.A}
\edef\spezttrra{\string{\csname\spzzttrra\endcsname\string}}
\iwritel\filefdef{\spezttrra}
\repeat  \immediate\closein\fileftmp \immediate\closeout\filefdef
\immediate\write16 {Panda Warning - fdef: Label(s) may have changed,
re-run TeX to get them right}\fi}
\def\iwritel#1#2{\newlinechar=-1
{\newlinechar=`\ \immediate\write#1{#2}}\newlinechar=-1}
\global\fdefcontre=0 \global\fdefcount=0 \global\indcount=0
%
%
\null
%
%
%
%

%
\loadamsmath
\loadeuler
\pageno=0
\nopagenumbers{\baselineskip=12pt
\line{\hfill CERN-TH-6758/92}
\line{\hfill\tt hep-th/9212100}
\line{\hfill December 1992}
\ifdoublepage \bjump\bjump\bjump\bjump\else\vfill\fi
\centerline{\capsone NEW TOPOLOGICAL THEORIES AND}
\sjump\sjump
\centerline{\capsone CONJUGACY CLASSES OF THE WEYL GROUP}
\bjump\bjump
\centerline{
{\scaps Timothy J. Hollowood\footnote{$^1$}{\tt hollow@surya11.cern.ch}}
and {\scaps J. Luis Miramontes\footnote{$^2$}{\tt miramont@cernvm.cern.ch}}}
\sjump
\sjump
\centerline{\sl Theory Division, CERN,}
\centerline{\sl CH-1211 Geneva 23, Switzerland}
\bjump\bjump\bjump
\ifdoublepage
\vfill
{\noindent
\line{CERN-TH-6758/92\hfill}
\line{December 1992\hfill}}
\eject\null\vfill\fi
\centerline{\capsone ABSTRACT}\sjump
The problem of interpreting a set of ${\cal W}$-algebra constraints
constructed in terms of an arbitrarily twisted scalar field as the
recursion relations of a topological theory is addressed. In this picture,
the conventional models of topological gravity coupled to $A$, $D$ or $E$
topological matter, correspond to taking the scalar field twisted by the
Coxeter element of the Weyl group.
It turns out that not all conjugacy classes of the Weyl group lead to a
topological model. For example, it is shown that for the
$A$ algebras there are two possible choices for the conjugacy class, giving
both the conventional and
a new series of topological models. Furthermore,
it is shown how the new series of theories
contains the conventional series as a subsector. A tentative interpretation
of this new series in terms of intersection theory is presented.
\sjump\vfill
\ifdoublepage \else
\noindent
\line{CERN-TH-6758/92\hfill}
\line{December 1992\hfill}\fi
\eject}
\yespagenumbers\pageno=1
%

\def\ss{{\eufm s}}

\def\W{$\cal W$}
\def\e{{\bfmath e}}
\def\vv{{\bfmath v}}
\def\M{\overline{\cal M}}
\mathchardef\bphi="731E
\mathchardef\balpha="710B
\def\pp{{\bfmath\bphi}}
\chapter{Introduction}

One of the consequences of the relationship between matrix
models and two-dimensional topological field theories is that
the coupling of two-dimensional topological gravity [\Ref{TOPG}] to
certain topological matter theories is described by the
solution of some integrable hierarchies
of equations of the Korteweg-de Vries (KdV) type, subject to an
additional constraint known as the {\it string equation} (see the review
[\Ref{REV}], and references therein). The flows of
the hierarchy are identified with the coupling constants to the operators
of the theory, and the partition function is the
tau-function of the hierarchy. Moreover, the solution is completely
specified by a set of \W-algebra constraints acting linearly on
the tau-function, which within the topological theory have a natural
interpretation in terms of the contact interactions of physical
operators.

For pure two-dimensional topological gravity, which is related to the
hermitian one matrix model, the hierarchy is precisely the
KdV hierarchy, the \W-constraints are simply Virasoro constraints, and
there is now a complete mathematical proof of the equivalence
[\Ref{KONT}].
In a similar fashion, it is supposed that the coupling of
two-dimensional topological gravity to the $A-D-E$ series of
topological minimal models is described by the Drinfel'd-Sokolov
$A-D-E$ generalized KdV hierarchies [\Ref{DS}],
and the \W-constraints [\Ref{WC}] are generated
by the classical \W-algebra associated to the Casimirs of the $A-D-E$
algebras [\Ref{WA}]. The $A_n$ case is also related to the
original hermitian $n$ matrix model [\Ref{DOUG}].

The Drinfel'd-Sokolov $A_n$ generalized KdV hierarchies can be
expressed in terms of (pseudo) differential scalar Lax operators
[\Ref{GD}], and
viewed as reductions of the Kadomtsev-Petviashvili (KP) hierarchy.
In fact, it was in this formalism that they were originally related to
the solutions of the matrix models [\Ref{DOUG}]. This formalism
has also provided a
quite explicit connection with the topological Landau-Ginzburg models
associated to the $A_n$ algebras [\Ref{REV}].
For the other algebras the situation is not quite as satisfactory.

One of the most surprising and beautiful aspects of the subject was developed
by Witten [\Ref{WIT}]. He showed how the correlation of topological gravity
could be interpreted as intersection numbers of certain bundles over the
moduli space of a punctured Riemann surface: in a sense the true topological
nature of the theory was revealed. This interpretation extends to
the case of $A$-type matter [\Ref{WIT},\Ref{MAT}].

{}From a more fundamental point of view, the generalized Drinfel'd-Sokolov
KdV hierarchies are associated to affine Kac-Moody algebras, and they
were originally defined in terms of {\it matrix} Lax operators (more
precisely gauge connections).
Recently, this construction has been generalized, and it has
been shown that it is possible to associate a hierarchy of the
KdV type to an affine Kac-Moody algebra and
a particular Heisenberg subalgebra.
The original Drinfel'd-Sokolov hierarchies
are then recovered by choosing the principal Heisenberg subalgebra.
Notice that in [\Ref{GEN1}], the construction
was undertaken in the loop algebra, whereas it was shown in [\Ref{GEN3}]
that these hierarchies are associated to the Kac-Moody algebras in
a representation independent way. Moreover, when the affine Kac-Moody
algebra is simply-laced, it has also been shown in [\Ref{GEN3}] that the
hierarchy is one of those constructed by Kac and Wakimoto within the
tau-function approach [\Ref{KW}].

The natural generalizations of the string equation  compatible with the
generalized Drinfel'd-Sokolov hierarchies have been obtained in [\Ref{PROC}].
When the hierarchy can be expressed using the tau-function formalism,
the result is that the tau-function corresponding to the solution of the
generalized string equation satisfies an infinite set of Virasoro
constraints, whose generators are constructed in terms of the elements
of the Heisenberg subalgebra. Similar results were obtained in [\Ref{VOS}]
directly in the tau-function formalism.
It is reasonable to suppose
that the tau-function will satisfy a complete set of \W-constraints
in a similar way to the case of the original Drinfel'd-Sokolov
hierarchies, where a complete proof is also lacking [\Ref{WCON}].

The simply-laced affine Kac-Moody algebras correspond to the untwisted
affinization of finite $A-D-E$ Lie algebras, and the consistency
with the Virasoro constraints requires that the generators of the
\W-constraints are always the generators of the classical \W-algebras
associated to the Casimirs of the corresponding finite Lie algebra
[\Ref{WA}].
These generators are constructed in terms of the elements of the
particular Heisenberg subalgebra that defines the hierarchy by a
generalized Sugawara construction.

Within this general framework, the natural question we address in the
following is whether there exist new topological points described by the
solutions of these new hierarchies, for which the
interpretation of the \W-constraints in terms of contact interactions
can be maintained. As yet there is no concrete definition of a topological
theory, but there are some generally accepted features that such a theory
could exhibit. First of all, at a ``topological point'' the
\W-constraints are recursion relations that allow one to calculate
an arbitrary correlation function at any genus exactly.
In tandem with this, there is a selection rule
which means that each correlation function
can be non-vanishing only on a Riemann surface with a specific number
of handles. This latter property is interpreted as the conservation of a
``ghost charge'' carried by the operators. The operators organize
themselves
into infinite sets, each set consisting of a unique ``primary'' and its
descendents, in such a way that any correlation function may be expressed
via the recursion relations in terms of correlation functions of the primaries
at genus zero.

A new topological model was discovered
in [\Ref{TAM}], and discussed more fully in [\Ref{PAS}], by looking at a
complex version of the non-linear Schr\"odinger hierarchy,
which is a different
hierarchy of the KdV-type associated to $A_1$.
At the level of the Virasoro constraints, the new model
corresponds to having a realization in terms of an untwisted scalar field,
as opposed to the twisted scalar field which describes conventional
topological gravity. The new topological model exhibits all of the features of
conventional topological gravity (except that it appears to have an operator
which has no descendants) in particular an intersection theory
interpretation of the correlation functions has been proposed which manifests
its topological credentials [\Ref{PAS}].

\chapter{General Formalism}

It is well known that the algebra
$A_1^{(1)}$ admits both a twisted and an untwisted representation in terms of
a scalar field. These two constructions lead
to pure topological gravity and the topological
model discussed at the end the introduction, respectively.
It is natural to ask whether
more general topological models exist for more general algebras.
Let us consider the untwisted affinization $g^{(1)}$ of a finite Lie algebra
$g$ of the $A$, $D$, or $E$ type. The central
object in the construction of generalized integrable hierarchies
is a particular Heisenberg subalgebra of
$g^{(1)}$, $\ss = {\Bbb C}\; c + \sum_{j\in E} {\Bbb C}\, b_j$, where
$E= I + {\Bbb Z} N$, with $I$ a set of rank($g$)
integers $\geq0$ and $<N$, for an integer $N$; the algebra being
$$
[b_j , b_k] = c\; {j\over N} \delta_{j+k,0}   \,,
\nfr{Hsa}
where $c$ is the central element of $g^{(1)}$. For each Heisenberg
subalgebra there is an associated gradation and a derivation
$d$ which counts the grade of the elements of $\ss$:
$$
[d, b_j] = {j\over N} b_j \,.
\nfr{Grade}
The non-equivalent Heisenberg subalgebras are classified
by the conjugacy classes of the Weyl group of the finite Lie algebra
$g$~[\Ref{KP}]. The
connection between a Weyl group element, say $w$ (up to conjugacy), and
the associated Heisenberg subalgebra $\ss_w$, is that there is a lift of
$w$ to $g^{(1)}$, denoted $\hat w$, which acts on the Heisenberg
subalgebra as
$$
\hat w (b_j) = {\rm exp} \left( {j\over N} 2\pi i\right) b_j\,.
\nfr{Endo}
In this case, $N$ is the order of $w$ whose eigenvalues are
${\exp}(2\pi i j/N)$, for $j\in I$. The original
Drinfel'd-Sokolov hierarchies are recovered by
considering only the principal Heisenberg subalgebra, for
which $w$ is the Coxeter element, $N$ is the Coxeter number, and $I$ is
the set of exponents of the algebra.

In the tau-function formalism, the hierarchy is constructed in terms of
a vertex operator representation of $g^{(1)}$ [\Ref{VERTEX}]
where the Heisenberg subalgebra is realized as
$$
c=1 \,,\quad  b_j = \cases{{\partial \over \partial t_j}\,;
& $j\geq0$\cr -{j\over N} t_{-j}\,; \quad & $j<0$\cr}\,,
\nfr{Fock}
acting on the Fock space ${\Bbb C}\;[t_j,\,\, j\in E>0] \otimes \{
{\rm exp} (\beta\cdot t_0),\,\, \beta\in \Lambda_0\}$, with $\Lambda_0$
being the root
lattice of $g$ projected onto the invariant subspace of $w$; $\Lambda_0$ is not
empty if $0\in I$.\note{In the following, we shall not label explicitly
any degeneracies that may arise in the set $I$ with the understanding
that in such cases one would have to introduce the appropriate
inner products in the
degenerate subspaces.} For the hierarchies of the KdV type that
we are considering, this vertex operator representation corresponds to
a {\it basic} level-one representation; hence, the hierarchy is defined
by the $t_j$-evolution of a tau-function
$$
\tau = \sum_{\beta\in\Lambda_0} \tau^{(\beta)} (t_j) {\rm e}^{\,\beta\cdot
t_0}\,;\quad j\in E>0\,.
\nfr{Tau}
The equations of the hierarchy follow from the condition that $\tau$
lies in the orbit of the highest weight state [\Ref{KW}].

The logarithm of the tau-function is the natural choice
for the ``free energy'' of the two-dimensional model
in analogy with the original
Drinfel'd-Sokolov hierarchies. Nevertheless, if
$0\in I$ then  $\Lambda_0$ is not empty, the tau function has an infinite
set of components $\tau^{\beta}$, $\beta\in\Lambda_0$, and the choice of the
``free energy'' does not seem to be unique. This is not the case for
the following reason. Notice that $b_0$
commutes with the other elements of $\ss_w$, therefore, acting on the
Fock space, it can be shifted arbitrarily along the invariant subspace of
$w$ without spoiling the algebra: $b_0 = \partial/\partial t_0 + \alpha$.
Hence, the eigenvalue of $b_0$ corresponding to each
$\tau^{(\beta)}$ is actually arbitrary, and all the components of the
tau-function are in fact equivalent
as candidates for the partition function. So we choose any one of them,
and the free energy will be\note{The
partition function of the $n$-matrix
model is actually the square of the tau-function.}
$$
F=\ln \tau (t,p)\,,
\nfr{Zf}
where the dependence on the (arbitrary) eigenvalue of $b_0$ is
explicitly indicated, $b_0 \tau(t,p)\equiv  p \tau(t,p)$.

The natural generalizations of the string equation for these
hierarchies have been discussed recently
in~[\Ref{PROC},\Ref{VOS}]. The result is that the tau-function \Tau\
satisfies an infinite set of Virasoro constraints
$$
L_{m} \tau(t,p) =0\,, \quad m\geq-1\,,
\nfr{Vircon}
where the Virasoro generators are constructed in terms of the elements of
$\ss_w$ as
$$
L_m = {1\over 2} \sum_{j+k = mN \atop j, k\in E} :b_j b_k:
+ \eta\,\delta_{m,0}\,, \quad
\eta={1\over 4N^2} \sum_{j\in I} j(N - j)\,;
\nfr{Suga}
the string equation is just the $L_{-1}$ constraint.
Acting on the Fock space, the Virasoro generators are
$$
\eqalign{
L_{-1} =& \sum_{j\in E> N} {j\over N} t_j {\partial\over \partial
t_{j-N}}
\,+\, {1\over 2 N^2} \sum_{ j\in I} j(N-j) t_j t_{N-j} \,+
\, pt_N\cr
L_0=& \sum_{j\in E>0} {j\over N} t_j{\partial\over\partial t_j} \,+\,
\eta\,+\,{1\over 2}p^2\cr
L_n=& \sum_{j\in E>0} {j\over N} t_j {\partial\over\partial t_{j+nN}}
\,+\, {1\over2}\sum_{0<j\in E<nN} {\partial^2\over\partial t_j
\partial t_{nN-j}} \,+\, p{\partial\over\partial t_{nN}}\,,\quad n>0
\cr}
\nfr{Virexp}
where $p=0$ if $0\notin I$.
Moreover, it is expected that the tau-function will
satisfy additional constraints, and
it has been conjectured in [\Ref{PROC}] that they should correspond
to $\cal W$-algebra constraints, a conjecture which is both natural and
the only possibility consistent with the flows of the hierarchy.
The consistency with the
Virasoro constraints requires that the relevant $\cal W$-algebra
corresponds to the Casimir algebra of $g$; hence, there will be a
generator $W^{(j+1)}$ for each exponent $j$ of $g$ that is constructed
in terms of $\ss_w$ by generalizing the Sugawara construction [\Ref{WA}].
So we are led to consider a model whose free energy is
\Zf, with $\tau$ constrained by $W^{(j+1)}_{i}\tau =0$,
$j$ being an exponent of $g$, and $i\geq -j$.

The interpretation of the integrable
hierarchy as a two-dimensional model is achieved through the
identification of the flows of the hierarchy with couplings to
(integrated) local operators. As usual, we have to shift the $t_j$'s
to go to the ``physical'' coupling constants
$$
t_j \rightarrow \beta \delta_{j,k} + t_j\,,
\nfr{Shift}
where $\beta$ is an arbitrary constant, and we have chosen some
$k\in E>0$. Then, the new $t_j$'s will be the coupling constants to
a certain set of operators $\{ O_j\,;\,\, j\in E>0\}$, such that
$$
\langle O_{j_1} \cdots O_{j_n} \rangle^{t} = {\partial\over \partial
t_{j_1}} \cdots {\partial\over \partial t_{j_n}} {\rm ln} \tau (t,p)\,;
\nfr{Corr}
$\langle\cdots\rangle ^t$
means that the correlation function depends on
$t_j,p\neq0$, and we shall omit the $^t$ when the correlation function
is considered at the point $t_j=p=0$.
If $0\in E$, $p$ will be understood as an additional coupling constant(s)
corresponding to another operator(s) $Q$ [\Ref{PAS}].

With these definitions, the $\cal W$-constraints become recursion
relations between correlation functions that allow to eliminate
certain operators. In particular, the Virasoro constraints can be
written as
$$
k\langle \beta O_{k+nN}\rangle^t= - {L_n \tau(t,p) \over \tau(t,p)}\,;
\quad n\geq-1\,,
\nfr{Elim}
and they can be used to eliminate the operators $O_{k+nN}$ with $n\geq-1$
from the correlation functions evaluated at $t_{j}=0$ for $j\geq k$.
Therefore, all these operators are {\it redundant}, and it makes
sense to consider the configuration where their coupling constants
vanish.
In general, notice that a generic \W-generator $W_{j}^{(r)}$ is a sum of
terms, each one being the normal-ordered product of $r$ elements of
$\ss_w$, $:b_{m_1} \cdots b_{m_r}:$ with $m_1 +m_2+\cdots+m_r =jN$.
Therefore, if the recursion relations provided by this constraint allow
for the elimination of a certain operator $O_n$ from the correlation functions
evaluated at $t_j=0$ for $j\geq k$, those provided by another
constraint $W_{l}^{(r)}$ will allow to eliminate the operator
$O_{n +(l-j)N}$.

If the model is topological, the \W-constraints should provide recursion
relations that allow one to calculate all these correlation
functions in a systematic algebraic fashion. According to the previous
discussion, this condition means that all the operators are {\it
redundant}, and then it makes sense to consider the configuration
where all the coupling constants vanish, which will be the ``topological
point''. To ensure this, it is enough to show that the set
of constraints $W^{(r)}_{1-r} \tau =0$ allow for the elimination of all the
operators $O_j$ with $j\leq N$ from the correlation functions.
The $L_{-1}$ constraint allows one to eliminate the operator $O_{k-N}$;
hence we choose $k=N+i$ with $i$ being the minimal element of $I>0$.
Then, the condition ensuring that the model is topological is that the
set of constraints $W_{1-r}^{(r)}\tau=0$ allows one to calculate all the
correlation functions of the
operators $O_j$, $j\leq N$, evaluated at $t_l=0$ for $l>N$. Notice that
the above argument implies that the solution of the constraints, up to
a trivial multiplicative factor, is unique in a neighbourhood of the
topological point.

Notice that the number of constraints equals the
number of operators $O_j$ with $j\leq N$, which is the rank of $g$. In
the topological model, these operators play the r\^ole of the ``primary
fields''. Of course, if $0\in I$, then $Q$ are additional
fields (there being more than one if $w$ has more than one
invariant direction), but their
correlation functions are also determined because
the \W-generators do not contain derivatives with respect to their
coupling constants $p$. The space spanned by the coupling constants of
these ``primary fields'', $t_j$=0 for $j\geq k=N+i$, will be called the
``small phase space'', in analogy with the terminology used in  the
conventional
topological field theories. The set of generators $W_{-j}^{(j+1)}$, $j$
being an exponent of $g$,
commute amongst themselves in the full coupling
constant space. Therefore, the condition that the model is topological is
just the condition that these generators also commute when restricted to
the small phase space. They can then be integrated and provide all
the correlation functions of the primary fields, which determine all
the other correlations functions of the model.

To connect our model with two-dimensional gravity, we have to
distinguish the contributions to the correlation functions corresponding
to Riemann surfaces with different number of handles.
We shall do it by introducing a genus counting parameter or ``string
coupling constant'', $\lambda$, such that
$$
\langle\cdots\rangle^t \equiv \sum_{h\geq 0} \lambda^{2(h-1)}
\langle\cdots\rangle^{t}_{h}\,.
\nfr{Stringcc}
The crucial question is how to introduce $\lambda$ in the equations of
the hierarchy. The relevant criterion is determined by preserving
the physical interpretation of the Virasoro constraints in the
conventional topological models, where they describe
the contact terms picked up when
one of the descendants of the puncture operator approaches either
another operator in the correlation function (a marked point on the
Riemann surface) or a node, and the Riemann surface factorizes
according to the stable compactification of its moduli space [\Ref{REV}].
Therefore, the
first term in $L_n$, $n>0$ \Virexp, should correspond to the process
when the
operator $\beta O_{k+nN}$ collides with one of the other operators
in the correlation function; under this factorization the genus $h$ of
the Riemann surface is not modified. The second term should be
associated to the process when $\beta O_{k+nN}$ approaches a node on the
surface. Then, either the genus of the Riemann surface
is lowered by one, if the node corresponds to pinching one handle,
or the Riemann surface is divided into two parts of genus $h'$ and
$h-h'$, if the node corresponds to a dividing cycle. To maintain this
interpretation of the Virasoro constraints,
all these terms should be present in the projection of the corresponding
recursion relations onto genus $h$. It is straightforward to see that
a relative factor of $\lambda^2$ is needed in front of the second
derivatives appearing in $L_n$. This interpretation forces the choice
$$
t_j \rightarrow \lambda^{\alpha j -1} t_j \,,\quad p\rightarrow
\lambda^{-1} p \,.
\nfr{Insert}
Notice that the Virasoro generators are homogeneous under the
transformation $t_j \rightarrow \lambda^{\alpha j} t_j$, {\it i.e.} they
transform as $L_n\rightarrow \lambda^{-\alpha nN}L_n$; therefore, the
value of $\alpha$ is arbitrary in principle.
Nevertheless, it will be fixed below by imposing the consistency of
\Insert, the Virasoro constraints, and the expansion \Stringcc\ at the
topological point. Within this interpretation,
the operator $O_i$ (where $i$ is the minimal element of $I>0$)
plays the r\^ole of the puncture-like operator with
descendants $O_{i+nN}$, $n>0$. We shall use the notation $P\equiv -\beta
O_i$ and $\sigma_n(P)\equiv -\beta O_{i+nN}$ in the following.

With these substitutions, the derivative $(\prod_{j\in A} \partial/
\partial
t_j) \partial^s/\partial p^s$ of \Elim, with $A$ being a set of indices
belonging to $E>0$, evaluated at $t_j=p=0$, and
projected onto genus $h$, provides the following recursion relations for
the correlation functions at the topological point\note{We have assumed
in the following
that $w$ has no more than one invariant direction, however,
generalizations are easily written down.}:
$$
\eqalignno{
{i+N\over N}\langle P & Q^s \prod_{j\in A} O_j\rangle_h =
\sum_{j\in A>N} {j\over N}
\langle Q^s \prod_{l\in A} O_{l - N\delta_{l,j}} \rangle_h
&\nameali{Puncture}\cr
{i+N\over N}\langle \sigma_1(P) & Q^s
\prod_{j\in A} O_j\rangle_h =
\left(\sum_{j\in A} {j\over N}\right) \langle
Q^s \prod_{l\in A} O_{l} \rangle_h
&\nameali{Scaling}\cr
{i+N\over N}\langle \sigma_{n+1}&(P) Q^s \prod_{j\in A} O_j\rangle_h =
\sum_{j\in A} j\langle Q^s \prod_{l\in A} O_{j+nN\delta_{l,j}}
\rangle_h + s\langle O_{nN} Q^{s-1} \prod_{j\in A} O_j\rangle_h & \cr
& +{1\over2} \sum_{0<j\in E<nN} \Biggl( \langle
O_j O_{nN-j} Q^s \prod_{l\in A} O_l\rangle_{h-1} & \cr
& -\sum_{{B\cup C=A \atop 0\leq r \leq s}\atop 0\leq
h'\leq h} \langle O_j Q^r \prod_{k\in B} O_k \rangle_{h'}
\langle O_{nN-j} Q^{s-r} \prod_{l\in C} O_l\rangle_{h-h'} \Biggr)\,,
&\nameali{Factor}\cr}
$$
where $n\geq1$ in \Factor, together with the following exceptions:
$$
\eqalignno{
& \langle P O_j O_l\rangle_{h=0} = {j l\over N(i+N)} \delta_{j+l, N}\,,
\qquad \langle P Q O_N\rangle_{h=0} = {N\over i+N} \,, &\nameali{ExcepA}
\cr
& \langle \sigma_1 (P) Q^2 \rangle_{h=0} = {N\over i+N} \,, \qquad
\langle \sigma_1(P)\rangle_{h=1} = {N \over i+N}\;\eta \,.
&\nameali{ExcepB} \cr}
$$
Notice that, if $0\notin I$, all the correlation functions
involving $Q^s$
with $s\neq 0$ should be ignored.
\Puncture\ is the analogue of the  ``puncture equation'', and the
different terms of \Factor\
represent the contact terms associated to the processes where
$\sigma_n(P)$ collides either with another operator or with a node
of the surface, as discussed before.

In addition to the $\cal W$-algebra constraints, \Insert\ provides a new
equation for the tau-function:
$$
\left[ \sum_{j\in E} (1-\alpha j) (t_j + \beta \delta_{j, i+N})
{\partial\over \partial t_j} + p{\partial\over\partial p} +
\lambda{\partial \over
\partial\lambda}\right] \tau(t,p) =0\,.
\nfr{Eqextra}
In terms of the correlation functions
at the topological point, \Eqextra\ means that
$$
[1- \alpha (i+N)] \langle \sigma_1(P) Q^s \prod_{j\in A} O_j
\rangle_h = \left[ \sum_{j\in A}(1-\alpha j) +  s + 2(h-1) \right]
\langle Q^s \prod_{j\in A} O_j \rangle_h\,.
\nfr{GhostA}
Now, the consistency between \GhostA\  and the exceptions of the
recurrence relations induced by the $L_0$ constraint \ExcepB\
dictates the value of the parameter $\alpha$:
$$
\alpha = {1\over i +N }\,.
\nfr{Alphval}
In addition, one finds that $F_h$ vanishes for $h\neq1$,
and $\langle Q^2\rangle_h$ vanishes for $h\neq 0$.

The recurrence relations \Scaling\ and
\GhostA, have a very important topological
interpretation. First, let us consider the sum of \GhostA\ with
\Scaling\ multiplied by $N/(i+N)$,
$$
\langle \sigma_1(P) Q^s \prod_{j\in A} O_j \rangle_h =
\left[ 2(h-1) + s + n(A) \right] \langle Q^s \prod_{j\in A} O_j\rangle_h
\,,
\nfr{Dileq}
where $n(A)$ is just the number of indices in the set $A$. We
discover that \Dileq\ is the analogue of the ``dilaton equation'',
showing that $\sigma_1(P)$ couples to the Euler characteristic of the
Riemann surface with $n(A)+ s$ marked points.

Finally, we can rewrite \GhostA\ as
$$
\left[ \sum_{j\in A} \left( {j-i\over N} -1\right)  - s\left(
{i\over N} + 1\right)
- 2{i+N\over N} (h-1) \right] \langle Q^s \prod_{j\in A} O_j \rangle_h
=0\,.
\nfr{Ghost}
This equation proves an important property of the model.
It means that a given correlation
function $\langle Q^s \prod_{j\in A} O_j \rangle_h$ vanishes
if $\sum_{j\in A} [j- (i+N)] - s(i+N) \not= 2(i+N)(h-1)$. This shows that
indeed each correlation function can only be non-vanishing for Riemann
surfaces whose number of handles $h$ is given by this {\it selection
rule}. Imitating the usual interpretation of this property within a
topological field theory, we introduce a conserved
``ghost number'' associated to each operator:
$$
q(O_j) = {j-i\over N} -1  \,,\quad q(Q)= -{i\over N} -1\,,
\nfr{Gnumber}
and the condition \Ghost\ can be expressed in the following way. The
correlation function of a set of operators $O_j$, $j\in U$, where now
$U$ can contain an arbitrary number of $Q$'s, can be non-vanishing
only on Riemann surfaces whose genus $h$ is given by the condition
$$
\sum_{j\in U} q(O_j) = 2{i+N\over N} (h-1) \,.
\nfr{Condition}
It is straightforward to check that all the exceptions, \ExcepA\ and
\ExcepB, are
consistent with this condition. Notice that if there is more than one $Q$
operator then they all have the same ghost number.
The assignment of ghost numbers \Gnumber\ indicates that the operators
$O_j$ with $0<j \in E\leq N$ play the r\^ole of ``primary operators''
in the topological model, while the other ones with $j>N$ correspond
to their descendants. In contrast, when $0\in I$, there is an additional
operator(s) $Q$ that does not have
descendants [\Ref{PAS}]. This is
related to the absence of derivatives with respect to its coupling
constant $p$ in the generators of the $\cal W$-algebra constraints.
It is worth noticing that the correlation functions involving only
primary
fields can be non-vanishing only at genus-zero. In fact, in the
topological model, the \W-constraints allow one to calculate any
arbitrary
correlation functions in terms of the correlation functions of the
primary operators at genus zero, which are themselves also determined by the
constraints. In this context, let us point out that the
\W-constraints do not fix the correlation functions involving only
the operator $Q$ because they do not contain derivatives with respect to
its coupling constant. Nevertheless, as a consequence of the selection
rule \Ghost, it is straightforward to realize that $\langle Q^s
\rangle_h$ has to vanish unless $s=2$ and $h=0$.

\chapter{Examples for $A$ algebras}

We now turn to explicit examples of the above formalism, in particular
we focus on the generalized hierarchies associated to $A_n$. The
Weyl group of $A_n$ is isomorphic to $S_{n+1}$,
the group of permutations of $n+1$ elements. The conjugacy classes of
$S_{n+1}$ are labelled by the possible partitions of $n+1$ into positive
integers:
$$
n+1 = n_1 + n_2 + \cdots+n_m\,, \qquad 1\leq n_m \leq \cdots
\leq n_1 \leq n+1\,.
\nfr{Partition}
The action of the Weyl group on the roots of $A_n$ can easily be determined
after the roots have been written in the standard way in terms of
a dependent set of $n+1$ vectors in the
root space of $A_n$ defined by\note{The simple roots
are ${\bfmath\balpha}_i=\e_i-\e_{i+1}$ for $i=1,2,\ldots,n$.}
$$
\sum_{j=1}^{n+1} \e_j =0 \,, \qquad \e_j\cdot \e_k = \delta_{j,k}
- {1\over n+1}\,.
\nfr{Vectors}
The conjugacy class corresponding to \Partition\ can then be
represented by the Weyl transformation that performs a cyclic permutation
within each set of vectors
$(\e_1,\ldots,\e_{n_1})$, $(\e_{n_1+1},
\ldots,\e_{n_1+n_2})$, $\ldots$ , $(\e_{n_1+\cdots+n_{m-1}+1},\ldots,
\e_{n+1})$.
The order $N$ of this Weyl transformation is the lowest common
multiple of $n_1,\ldots,n_m$, and the eigenvalues are
$$
I=\{\underbrace{0,\ldots,0}_{ m-1 \;{\rm times}}\}
\bigcup_{j=1}^{m}
\left\{{N\over n_j}, 2{N\over n_j},\ldots,(n_j -1){N\over n_j}\right\}\,.
\nfr{Indices}

The generators of the classical \W--algebra associated to
the Casimirs of $A_n$ are also constructed in terms of the
set of vectors \Vectors. They are given by~[\Ref{WGEN}]
$$
W^{(k)}(z)=\sum_{J_k}:\prod_{j_a\in J_k}\e_{j_a}\cdot\partial{\pp}(z):
\nfr{Wexpre}
where $J_k$ is a subset of $k$ different
integers between $1$ and $n+1$ (regardless of
order), and the sum is over the $\left({n+1\atop k}\right)$ such subsets.
The modes of the generators are
$$
W^{(k)}(z) = \sum_{j\in{\Bbb Z}} z^{-j-k} W^{(k)}_{j}\,.
\nfr{Wgen}
The scalar field $\partial \pp(z)$, twisted by the element of the Weyl group
$w$, is constructed in terms of the elements of $\ss_w$ as
$$
\partial\pp(z) = \sum_{j\in I} \vv (j)\partial\phi_j(z)\,, \quad
\partial\phi_j(z) =\sum_{k\in{\Bbb Z}} z^{-{j\over N}
-k-1} b_{j+kN}\,,
\nfr{Field}
where $\vv(j)$ is an  eigenvector of $w$ corresponding to the
eigenvalue ${\rm exp} (2\pi ij/N)$ normalized as
$$
\vv(j)\cdot\vv(k) = \delta_{j+k,0\;{\rm mod}\,N}\,,\quad j,k\in I\,.
\nfr{Normv}
Notice that we have not allowed for any Feign-Fuchs-like
deformation of the \W-algebra generators because such a deformation is
already accounted for by the arbitrary value of the zero-mode $p$.

We are now in position to discuss the possible topological points
corresponding to a tau-function that satisfies the $\cal W$-algebra
constraints
$$
W_{j}^{(k)} \tau =0\,, \quad k = 2,\ldots, n+1\,, \quad j\geq 1-k\,.
\nfr{Ancase}
According to our previous general discussion, the condition
for having a topological
configuration is that the constraints
$$
W_{1-k}^{(k)}\tau =0\,,\quad k=2,\ldots,n+1\,,
\nfr{Primary}
allow one to obtain the free energy in the small phase space, spanned by
the coupling constants
of the primary operators: $p$ and $t_j$ with $0<j\in E\leq N$.
Recall that the  $\cal W$-generators of the
constraints \Primary\ commute among
themselves in the full Fock space. Therefore, it will be possible to
integrate them if, when restricted to the small phase space, they do
not contain derivatives with respect to any coupling outside
the small phase space. From this perspective
the most ``dangerous'' term in $W^{(k)}_{1-k}$ is
$$
b_{-j_1}b_{-j_2}\cdots b_{-j_{k-1}}b_l\sim
\prod_{a=1}^{k-1}\left(t_{j_a}+\beta\delta_{j_a,N+i}\right)
{\partial\over\partial t_l},
\efr
with $j_1,\ldots,j_{k-1}
\geq0$, $l>0$ and $l=(1-k)N + j_1+\cdots+j_{k-1}$; therefore,
$l\leq (k-1)i=(k-1)N/n_1$ (recall that $n_1\geq n_2\geq\cdots \geq
n_m$ in \Partition). Notice that, \Indices\ implies that
$(k-1)N/n_1\in E$, hence, in the small phase space, the generator
$W_{1-k}^{(k)}$ may contain derivatives with respect to $t_l$ with
$l\leq (k-1)N/n_1$. The condition for the integrability of these
constraints in the small phase space is therefore $(k-1)N/n_1 \leq N$ for
$k=2,\ldots, n+1$: in short, $n\leq n_1$. This condition is only
satisfied by two partitions. The first one is that with $m=1$ and
$n_1=n+1$,
which corresponds to the conjugacy class of the Coxeter element of
the Weyl group of $A_n$, and the second one has $m=2$, $n_1=n$ and
$n_2=1$.
In these two cases, the set of generators of the constraints
\Ancase\ are first order differential operators with respect to the
coupling constants of the small phase space, and they indeed allow one to
obtain all the correlation functions of the primary operators.

Therefore, we conclude that, within this formalism,
only two different topological models
are described by a generalized $A_n$ hierarchy.
The first one corresponds to the hierarchy
associated to the conjugacy class of the Weyl group including the
Coxeter element, {\it i.e.} the principal Heisenberg subalgebra,
for which $N=n+1$ and $I=\{1,2,\ldots,n\}$. This is, of course, nothing
but the well-known model arising from
the hermitian n-matrix model and two-dimensional topological
gravity coupled to the twisted $A_n$ minimal $N=2$ SCFT.

The new topological model corresponds to the conjugacy class of the Weyl
group of $A_n$ specified by the partition $n+1 = (n)+(1)$. Therefore,
$N=n$, $I=\{0,1,\ldots,n-1\}$, and the set of ``primary
fields'' is $\{O_0\equiv Q,O_1,\ldots,O_n\}$. It follows from \Gnumber\ that
the ghost numbers of the operators in this model are
$$
q(O_j) = {j-1\over n} -1\,, \qquad j\geq0\,,
\nfr{GAn}
and the correlation function of a given set of operators
on a Riemann surface with $h$ handles has to vanish  unless
the sum  of their ghost numbers satisfies
$$
\sum q(O_j) = 2\;{n+1\over n}\;(h-1)\,.
\nfr{Gnew}
This should be compared with the case of the model associated to the
principal Heisenberg subalgebra, where $N=n+1$,
the primary fields are $O_1,
\ldots,O_n$, the ghost numbers are $q(O_j)=(j-1)/(n+1) -1$, and their
correlation functions vanish unless the sum of the ghost
numbers is $2(n+2)(h-1)/(n+1)$.

For completeness, we include the explicit form of the free energy
in the small phase space that follows from integration of the
constraints \Ancase\ corresponding to
the lowest values of $n$. For $A_1$, the
new topological model is a solution to the hierarchy associated of
the homogeneous gradation, {\it i.e.\/} the non-linear Schr\"odinger
hierarchy. Therefore this is precisely the model discovered
within the
context of hermitian one matrix models in the two arc sector~[\Ref{TAM}],
and discussed in more detail in~[\Ref{PAS}]. Its free energy in the small
phase space is
$$
\lambda^2 {\rm ln} \tau = -{1\over 4\beta} p t_{1}^{2} \,.
\nfr{Nls}
For $A_2$, the new topological model is associated to the ``intermediate''
hierarchy considered in~[\Ref{GEN1},\Ref{VD}], and its free energy is
$$
\lambda^2 {\rm ln} \tau = -{1\over 36\beta} t_{1}^3 - {2\over
3\beta} p \left( t_{1}t_{2} -{2\over 3\beta} t_{2}^{3}\right)\,.
\nfr{Atwo}
In a similar fashion, the free energy for the new topological model
associated to $A_3$ is
$$
\lambda^2 {\rm ln} \tau  = - {1\over 12\beta} t_{1}^2 t_{2}
- {1\over 72\beta^2} t_{2}^{4}
-{3\over 4\beta} p \bigg( t_{1} t_{3} -{1\over3} t_{2}^2
 +{3\over2\beta} t_{2}t_{3}^{2}
- {9 \over  8\beta^2}   t_{3}^{4} \bigg)\,.
\nfr{Athree}
Notice that, as follows from the discussion at the end of the previous
section, one could add an arbitrary term proportional to $p^2$ to these
expressions.

The ghost number selection rule \Gnew\ suggests that there is a
relationship between the new topological model associated to $A_n$ and
the conventional topological model associated to $A_{n-1}$. Indeed, the
Weyl group element used to define the new models acts as the Coxeter element
on a subalgebra $A_{n-1}\subset A_n$, but has, in addition, an invariant
direction. It turns out, as we show below, that the partition
function of the new topological model associated to $A_n$, for $n\geq2$,
reduces to the partition function of the conventional topological model
associated to $A_{n-1}$ when $p=0$. Notice by inspection that this is
indeed the case in \Atwo\ and \Athree.

In order to prove this relationship, let us
explicitly consider the contribution of the untwisted component of the
scalar field, $\partial\phi_0(z)$, to the generators of the
\W-constraints of the new model. Taking into account
that $\partial\phi_0(z)$ commutes with all the other components of $\partial
\pp(z)$, it follows that
$$
W_j^{(k)} = \sum_{l=0}^{k} C_{l}^{(k)}\sum_{m+r=j}
:\left[\partial\phi_0( z)^{k-l} \right]_m :{\widetilde W}_r^{(l)} \,,
\nfr{Relation}
where $[\partial\phi_0(z)^k]_m$ is just the coefficient of
$z^{-m-k}$ in $\partial\phi_0(z)^k$, and ${\widetilde W}_{r}^{(l)}$
is one of the \W-constraint generators for the Coxeter-twisted
$A_{n-1}$ subalgebra. In addition, $C^{(k)}_l$ are constant coefficients with
$C_{k}^{(k)}=1$, for $k=1,2,\ldots, n$, and $C_{n+1}^{(n+1)}=0$.
We denote the partition function of the conventional topological model
associated to $A_{n-1}$ as ${\widetilde\tau}$: it is a function only of
$t_{j+mn}$ with $j=1,2,\ldots,n-1$ and $m\in{\Bbb Z}\geq0$, which we denote
collectively as $\tilde t$. Taking $p=0$ in \Relation,
we see that $W_{j}^{(k)}{\widetilde\tau}$ is only non-vanishing
when $m\leq l-k$ and $r\leq -l$; hence if
$j\leq-k$. But the constraints are only defined for
$j>-k$ and so this proves that ${\widetilde\tau}$ is a solution of the
\W-constraints \Ancase\ at $p=0$. We now appeal to the uniqueness of the
solution to the \W-constraints which implies that $\tau$ is in fact equal
(with a suitable normalization) to $\widetilde\tau$ at $p=0$.
Consequently, the free energy of the new topological model associated to
$A_n$ can be written as
$$
{\rm ln}\tau (t,p) = {\rm ln}{\widetilde\tau}
(\tilde t) + p f(t,p)\,,
\nfr{Reduce}
where ${\rm ln}{\widetilde\tau}(\tilde t)$ is the free energy of the
conventional topological model associated to $A_{n-1}$.

The relation \Reduce\ has two direct consequences for the correlation
functions. Firstly, the correlation functions involving only operators
$O_{j+mn}$ with $j=1,2,\ldots,n-1$ and $m\in{\Bbb Z}\geq0$, coincide with
those of the conventional model associated to
$A_{n-1}$. Moreover, the correlation functions involving at least
one operator $O_{mn}$ with $m\in{\Bbb Z}>0$, but not the operator
$Q$, vanish.

For the original $A_n$ topological models, the models for $n>1$ can
all be understood as the addition of topological matter fields
$O_j$, for $j=2,3,\ldots,n$ to the basic theory of topological gravity
which only has one primary operator --- the puncture operator $P\simeq O_1$.
For the new series of topological models, it appears that the theories
associated to $A_n$ can be thought of as the new $A_1$ theory, with
primaries $Q$ and $O_n$, coupled to a set of matter fields
$O_j$, for $j=2,3,\ldots,n-1$ and the puncture operator of the conventional
theory $O_1$; however, $Q$ has no descendants.

\chapter{Discussion}

In~[\Ref{PAS}], a possible mathematical description
of the topological
model corresponding to the NLS hierarchy was proposed in terms of an
intersection theory on Riemann surfaces along the lines of~[\Ref{WIT}].
We now show that there is a generalization of this conjecture to the
whole new $A_n$ series, in the similar way that the conventional $A_n$
series can be understood [\Ref{REV},\Ref{WIT},\Ref{MAT}]. First of all, the
relation \Reduce\ provides a strong clue in the search for the correct
interpretation in terms of intersection theory, as we shall see.

Sticking to the spirit of [\Ref{WIT},\Ref{MAT}] and with
[\Ref{PAS}] in mind, the generalization proceeds as
follows. Consider a Riemann surface $\Sigma$ with $h$ handles and
$r+s$ marked points, $x_1,\ldots,x_r,y_1,\ldots,y_s$,
and its Deligne-Mumford compactified moduli space
$\M_{h,r+s}$. Label each $x_j$ with an integer
$0\leq k_j\leq n-1$, and consider the line bundle
$$
{\cal S} = {\cal K}^{n-1} \bigotimes_{j=1}^{r} {\cal O}(x_j)^{k_j}
\bigotimes_{l=1}^{s} {\cal O}(y_l)^{-1}
\nfr{Bundle}
over $\M_{h,r+ s}$, where $\cal K$ is the canonical line bundle of
$\Sigma$. So the sections of
$\cal S$ are $n-1$ forms and there exists a section with
simple zeros at the points $y_l$ and
poles of order $k_j$ at the points $x_j$, being otherwise regular.
The degree of $\cal S$ is ${\rm deg}({\cal S}) =
2(n-1)(h-1) + \sum k_j -s$. If this degree is a multiple of $n$, we
can define another line bundle $\cal T$ as the $n^{\rm th}$ root of
$\cal S$. Actually, there are $n^{2h}$ choices, and this defines a
branched covering of $\M_{h,r+s}$, that is denoted as
$\M_{h,r+s}'$. Now, let us consider the vector bundle $\cal V$ over the
moduli space whose fiber ${\cal V}_\Sigma$
at a point $\Sigma\in {\cal M}_{h,r+s}$ is
defined as the space of holomorphic sections of $\cal T$:
${\cal V}_\Sigma = H^0(\Sigma,{\cal T})$, if $H^1(\Sigma,{\cal T})=0$
(see [\Ref{WIT}] for a more precise definition of ${\cal V}$ which
takes account of the possibility that $H^1(\Sigma,{\cal T})$ may not
vanish). This vector bundle $\cal V$ has Chern classes and
the complex dimension of the top dimensional Chern class
$c_D({\cal V})$ can be found by using the Riemann-Roch theorem:
$$
D = {n-2\over n  } (h-1) + \sum_{j=1}^r {k_j\over n} -{s\over n}\,.
\nfr{Dimension}
Of course $c_D({\cal V})=0$ if $D\leq0$.
The last step is to consider, for each $x_j$, the complex line bundle
${\cal L}_j$ whose fiber is the cotangent bundle to $\Sigma$ at $x_j$,
and its first Chern class $c_1({\cal L}_j)$. Now, we
associate with each marked point $x_j$ a non-negative integer $m_j$,
in addition to the $k_j$'s, $0\leq k_j\leq n-1$, introduced earlier,
and we symbolically let $\sigma_{m_j}(j)=c_1({\cal L}_j)^{m_j}$. The
generalization of the conjecture in~[\Ref{PAS}] would be that the
correlation functions of the new topological model associated to $A_n$
are the following intersection numbers
$$
\left\langle Q^s \;\prod_{j=1}^{r} O_{k_j + 1 +nm_j } \right\rangle_h
{\buildrel ? \over =}
\left(\sigma_{m_1}(1)\cdots \sigma_{m_r}(r) \cdot c_D({\cal V})
\,,\, \M_{h,r+s}'\right)\,.
\nfr{Conjecture}
Of course, these numbers vanish unless a certain dimensional condition
is satisfied. In particular, the complex dimension of the product of
cohomology classes has to equal the complex dimension of
$\M_{h,r+s}'$:
$$
\sum_{j=1}^r m_j + D = {\rm dim}\left( \M_{h,r+s}' \right)
= 3(h-1) + r+ s \,,
\nfr{Dimcond}
with $D$ given by \Dimension. This condition is precisely the ghost
number selection rule \Gnew\ for the correlation functions of the new
series of topological models associated the $A$-algebras.
Of course, as has already been emphasized
in~[\Ref{PAS}], this is just the first tentative step in proving that the
topological model does indeed generate the proposed intersection numbers.

In support of the conjecture we can say the following:
suppose that we now consider a correlation function which has
no insertions of $Q$ and no insertions of $O_{mn}$, for $m\in{\Bbb Z}\geq1$.
That means that we consider the above construction with $s=0$ and $k_j<
n-1$ and hence the
construction reduces to that for the $A_{n-1}$
conventional topological models in [\Ref{WIT},\Ref{MAT}]. But this was
anticipated in \Reduce, which shows that with no insertions of $Q$ the
new $A_n$ model reduces to the conventional $A_{n-1}$ model. Actually,
\Reduce\ is stronger since it implies that any correlation function
with insertions of $O_{mn}$, $m\in{\Bbb Z}\geq1$, but with no insertions
of $Q$ vanishes. This is something which should be understood in the
context of intersection theory.

Summarizing, we have shown that the $A$ algebras admit two distinct
series of topological models, one of which is the well-known series
describing topological gravity coupled to topological $A$-type matter and
the other is a new series of models of
which the first is that of [\Ref{TAM},\Ref{PAS}].
Furthermore, we have shown that the conventional model of topological
gravity coupled to $A_n$ topological matter is contained in a subsector
of the new $A_{n+1}$ topological model. At the moment the physical
meaning of the new models and their relation to the conventional models
is somewhat obscure. In particular, one would like to have a physical
understanding of the model of [\Ref{TAM},\Ref{PAS}], the
new $A_1$ model, since the generalizations discussed in this papers
can be thought
of as the coupling of it to the conventional models. Additionally, it would be
interesting to consider the other algebras in a similar way.

\references

\beginref
\Rref{TOPG}{J.M.F. Labastida, M. Pernici and E. Witten, Nucl. Phys.
{\bf B310} (1988) 611 \newline
D. Montano and J. Sonnenschein, Nucl. Phys. {\bf B313} (1989) 258;
Nucl. Phys {\bf B324} (1990) 348 \newline
R. Myers and V. Periwal, Nucl. Phys. {\bf B333} (1990) 536}
\Rref{REV}{R. Dijkgraaf, {\sl Intersection Theory, Integrable Hierarchies
and Topological Field Theory\/}, IAS preprint IASSNS-HEP-91/91,
{\tt hep-th/9201003}, {\it to appear in the Proceedings of the Carg\`ese
Summer School on New Symmetry Principles in Quantum Field Theory,
(1991)}}
\Rref{KONT}{M. Kontsevich, Commun. Math. Phys. {\bf147} (1992) 1\newline
E. Witten, {\sl On the Kontsevich model and other models of two
dimensional gravity}, IAS preprint IASSNS-HEP-91/24 \newline
D.J. Gross and M.J. Newman, Nucl. Phys. {\bf B380} (1992) 168}
\Rref{DOUG}{M. Douglas, Phys. Lett. {\bf B238} (1990) 176}
\Rref{GD}{I.M. Gelfand and L.A. Dickey, Funct. Anal. Appl. {\bf11} (1977)
93\newline
L.A. Dickey, {\sl Soliton equations and Hamiltonian systems\/}, Adv. Ser.
Math. Phys. Vol. {\bf12}, World Scientific 1991}
\Rref{WA}{F.A. Bais, P. Bouwknegt, K. Schoutens and M. Surridge, Nucl.
Phys. {\bf B304} (1988) 348; Nucl. Phys. {\bf B304} (1988) 371}
\Rref{WC}{K. Li, Nucl. Phys. {\bf B354} (1991) 711; Nucl. Phys. {\bf
B354} (1991) 725}
\Rref{WIT}{E. Witten, Surv. in Diff. Geom. {\bf 1} (1991) 243;
{\sl Algebraic geometry associated with matrix models of two dimensional
gravity}, IAS preprint IASSNS-HEP-91/74}
\Rref{MAT}{E. Witten, Nucl. Phys. {\bf B371} (1992) 191}
\Rref{GEN1}{M.F. de Groot, T.J. Hollowood and J.L. Miramontes,
Commun. Math. Phys. {\bf145} (1992) 57\newline
N.J. Burroughs, M.F. de Groot, T.J. Hollowood and J.L.
Miramontes, Phys. Lett. {\bf B277} (1992) 89}
\Rref{GEN3}{T.J. Hollowood and J.L. Miramontes, {\sl Tau functions
and generalized integrable hierarchies\/}, Oxford preprint
OUTP-92-15P, CERN preprint CERN-TH-6594/92, {\tt hep-th/9208058},
{\it to appear in Commun. Math. Phys.}}
\Rref{PROC}{T.J. Hollowood, J.L. Miramontes and J. S\'anchez Guill\'en,
{\sl Generalized integrability and two-dimensional gravitation\/},
CERN preprint CERN-TH-6678/92, {\tt hep-th/9210066}, {\it to appear in
the Proceedings of the Conference on Modern Problems in Quantum Field
Theory, Strings and Quantum Gravity, Kiev (1992)}}
\Rref{KP}{V.G. Kac and D.H. Peterson, {\sl 112 constructions of the
basic representation of the loop group of $E_8$\/}, in: Symp. on
Anomalies, geometry and topology, eds. W.A. Bardeen and A.R. White.
World Scientific, Singapore, 1985\newline
V.G. Kac, {\sl Infinite dimensional Lie algebras\/}, $3^{\rm rd}$ edition,
Cambridge University Press 1990}
\Rref{KW}{V.G. Kac and M. Wakimoto, Proc. of Symp. in Pure
Math. {\bf 49} (1989) 191}
\Rref{VOS}{K. de Vos, Nucl. Phys. {\bf B375} (1992) 478}
\Rref{DS}{V.G. Drinfel'd and V.V. Sokolov, J. Sov. Math. {\bf30}
(1985) 1975}
\Rref{TAM}{\v C. Crnkovi\'c, M. Douglas and G. Moore, Nucl. Phys. {\bf B360}
(1992) 507}
\Rref{PAS}{A. Pasquinucci, Mod. Phys. Lett. {\bf A7} (1992) 2299}
\Rref{WCON}{R. Dijkgraaf, E. Verlinde and H. Verlinde,
Nucl. Phys. {\bf B348} (1991) 435\newline
M. Fukuma, H. Kawai and R. Nakayama,
Int. J. Mod. Phys. {\bf A6} (1991) 507; Commun. Math. Phys. {\bf143}
(1992) 403\newline
V.G. Kac and A. Schwarz, Phys. Lett. {\bf B257} (1991)
329\newline
J. Goeree, Nucl. Phys. (1991) 737\newline
S. Panda and S. Roy, {\sl The Lax operator approach for the Virasoro
and the W-constraints in the generalized KdV hierarchy},
ICTP preprint IC-92-226, {\tt hep-th/9208065}}
\Rref{VERTEX}{M.B. Halpern, Phys. Rev. {\bf D12} (1975) 1684;
Phys. Rev. {\bf D13} (1976) 337 \newline
I.B. Frenkel and V.G. Kac, Invent. Math. {\bf62}
(1980) 23\newline
J. Lepowsky and R.L. Wilson, Commun. Math. Phys. {\bf62} (1978) 43
\newline
V.G. Kac, D.A. Kazhdan, J. Lepowsky and R.L. Wilson, Adv. in
Math. {\bf42} (1981) 83\newline
J. Lepowsky, Proc. Natl. Acad. Sci. USA {\bf82} (1985) 8295}
\Rref{WGEN}{V.A. Fateev and S.L. Lykyanov, Int. J. Mod. Phys. {\bf A3}
(1988) 507\newline
V.A. Fateev and A.B. Zamolodchikov, Nucl. Phys. {\bf B280}
[FS18] (1987) 644}
\Rref{VD}{P. van Driel, Phys. Lett. {\bf B274} (1992) 179}
\endref
\ciao
